\documentclass{PoS}

\title{Review on low and high mass spectroscopy}

\ShortTitle{Review on low and high mass spectroscopy}

\author{\speaker{Chang-Zheng Yuan}\\
Institute of High Energy Physics, Chinese Academy of
Sciences, Beijing 100049, China\\
E-mail: \email{yuancz@ihep.ac.cn}}

\abstract{We review the most recent experimental progress in the
hadron spectroscopy, up to bottomonium mass. This covers the
search for the $J^{PC}=1^{-+}$ exotic state, the states in
$J/\psi$ decays, the spin-singlets of heavy quarkonium, and the
charmoniumlike $XYZ$ states. }

\FullConference{35th International Conference of High
Energy Physics - ICHEP2010,\\
July 22-28, 2010\\
Paris France}

\begin{document}

\section{Introduction} 

In the quark model, mesons are the bound state of one quark and
one anti-quark, while baryons are composed of three quarks.
Although no solid calculation shows hadronic states with other
configurations must exist in QCD, people believe hadrons with no
quark (glueball), with excited gluon (hybrid), or with more than
three quarks (multi-quark state) exist. Since a proton and a
neutron can be bounded to form a deuteron, it is also believed
that other mesons can also be bounded to form molecules.

It is a long history of searching for all these kinds of states,
however, no solid conclusion was reached until now on the
existence of any one of them, except deuteron. In this talk, we
show the most recent progress in the experimental study of the
light and heavy hadrons.

\section{The state with exotic quantum number}

States with exotic quantum numbers that quark and antiquark pair
can not have give clear evidence of the existence of
non-$q\bar{q}$ states. So far, three candidates with
$J^{PC}=1^{-+}$ were reported but some of them suffer from
controversial experimental information. A partial wave analysis of
2004 COMPASS data~\cite{compass} from the diffractive dissociation
of 190~GeV $\pi^-$ on a lead target into the $\pi^-\pi^-\pi^+$
final state was performed. In addition to well-known $q\bar{q}$
states, a significant spin-exotic wave with $J^{PC}=1^{-+}$
decaying to $\rho\pi$ is found. The mass and width are $1660\pm
10^{+0}_{-64}$~MeV and $269\pm 21^{+42}_{-64}$~MeV, respectively.
Its mass-dependent phase differences to the $J^{PC}=2^{-+}$ and
$1^{++}$ waves are consistent with the highly debated
$\pi_1(1600)$ meson. There are 200 times more events in COMPASS
2008 data, more results on light hadron spectroscopy are expected
to come soon.

\section{States in $J/\psi$ decays}

With newly accumulated 106~M $\psi(2S)$ and 226~M $J/\psi$ events,
BESIII experiment~\cite{huangyp} confirmed the existence of the
threshold enhancement in $J/\psi\to \gamma p\bar{p}$ final state,
with mass agrees with the BESII measurement, and the width less
than 8~MeV at the 90\% confidence level (C.L.). In an analysis of
the $J/\psi\to \gamma \eta^\prime \pi^+\pi^-$, three resonances
are observed. The $X(1835)$ has a mass of $1838\pm 3$~MeV, in good
agreement with BESII result, and a width of $180\pm 9$~MeV, which
is larger than the BESII measurement. The two new structures are
at 2124~MeV and 2371~MeV, with widthes around 100~MeV. In the
$\eta\pi^+\pi^-$ invariant mass recoiling against an $\omega$ in
$J/\psi$ decays, besides the known $f_1(1285)$ and $\eta(1405)$, a
state at $1873\pm 11$~MeV with a width of $82\pm 19$~MeV is seen,
it could be the hadronic production of the $X(1835)$ observed in
$\eta^\prime \pi^+\pi^-$ mode, although the mass difference is
large. The nature of these states is unknown, with the
possibilities of being the excited $\eta$ or $\eta^\prime$ states,
the glueballs, the $p\bar{p}$ molecular states, and so on.

\section{Spin-singlet charmonium and bottomonium states}

The $P$-wave charmonium spin-singlet state $h_c$ is studied at
BESIII with 106~M $\psi(2S)$ events accumulated in
2009~\cite{bes3hc,ligang}. Clear signals are observed for
$\psi(2S)\to \pi^0 h_c$ with and without the subsequent radiative
decay $h_c\to \gamma\eta_c$. First measurements of the absolute
branching ratios $\mathcal{B}(\psi(2S) \to \pi^0 h_c) = (8.4 \pm
1.3 \pm 1.0) \times 10^{-4}$ and $\mathcal{B}(h_c \to \gamma
\eta_c) = (54.3 \pm 6.7 \pm 5.2)\%$ are determined.  A statistics
limited determination of the previously unmeasured $h_c$ width
leads to an upper limit $\Gamma(h_c)<1.44$~MeV at the 90\% C.L.
Measurements of $M(h_c) = 3525.40 \pm 0.13 \pm 0.18$~MeV and
$\mathcal{B}(\psi' \rightarrow \pi^0 h_c) \times \mathcal{B}(h_c
\rightarrow \gamma \eta_c) = (4.58 \pm 0.40 \pm 0.50) \times
10^{-4}$ are consistent with previous results by
CLEOc~\cite{pdg2010}.

Both BaBar~\cite{Druzhinin} and Belle~\cite{Nakazawa} experiments
found new decay modes of the radial excitation of the $S$-wave
charmonium spin-singlet, the $\eta_c(2S)$, and measured the mass
and width of it. The new decay modes are $K^+K^-\pi^+\pi^-\pi^0$,
$K_SK^+\pi^+\pi^-\pi^-+c.c.$, $K^+K^-2(\pi^+\pi^-)$, and
$3(\pi^+\pi^-)$, these add to the previous known modes
$K\bar{K}\pi$ and $\gamma\gamma$. The masses are measured to be
$3638.3\pm 1.5\pm 0.5$~MeV (BaBar) and $3636.9\pm 1.1\pm 2.5\pm
5.0$~MeV (Belle), and the widths are $14.2\pm 4.4\pm 2.5$~MeV
(BaBar) and $9.9\pm 3.2\pm 2.6\pm 2.0$~MeV (Belle). Together with
the previous measurements~\cite{pdg2010}, we get the best
estimation of the mass and width of the $\eta_c(2S)$ to be
$M=3638.0\pm 1.4$~MeV and $\Gamma=12.2\pm 3.1$~MeV.

The bottomonium $S$-wave spin-singlet $\eta_b$ was observed in
$\Upsilon(2S)$ and $\Upsilon(3S)$ radiative transitions by
BaBar~\cite{babar_etab} and CLEO~\cite{cleo_etab} experiments. The
transition rates are at a few $10^{-4}$ level and the mass
splitting from the spin-triplet is $69.3\pm 2.8$~MeV. Most recent
theoretical calculation gives consistent result but with large
uncertainty: $60.3\pm 5.5(\rm stat.)\pm 3.8(\rm sys.)\pm 2.1(\rm
exp.)$~MeV~\cite{Meinel}. The calculation predicts the mass
splitting of the $2S$ states of $23.5\pm 4.1\pm 1.5\pm 0.8$~MeV,
however, the observation of $\eta_b(2S)$ should be extremely hard
due to the low energy of the radiative photon from $\Upsilon(2S)$
or $\Upsilon(3S)$ decays.

BaBar searched for the $P$-wave bottomonium spin-singlet $h_b$ in
$\Upsilon(3S)\to \pi^+\pi^-h_b$ and $\pi^0h_b$~\cite{fulsom}. No
evidence was found in $\pi^+\pi^-h_b$ mode, and ${\cal
B}(\Upsilon(3S)\to \pi^+\pi^-h_b)<1\times 10^{-4}$ was determined
at the 90\% C.L. There is a faint evidence (2.7$\sigma$) at
9903~MeV in $\Upsilon(3S)\to \pi^0h_b$, and the production rate is
determined to be ${\cal B}(\Upsilon(3S)\to \pi^0h_b)=(3.1\pm
1.1\pm 0.4)\times 10^{-4}$.

\section{The charmoniumlike $XYZ$ states}

As the $B$-factories accumulate more and more data, lots of new
states have been observed in the final states with a charmonium
and some light hadrons. All these states populate in the
charmonium mass region. They could be candidates for usual
charmonium states, however, there are also lots of strange
properties shown from these states, these make them more like
exotic states rather than conventional charmonium
states~\cite{qwg_review,pic2009}.

\subsection{The $X(3872)$}

The $X(3872)$ was discovered by Belle in 2003~\cite{belle_x3872}
as a narrow peak in the $\pi^+\pi^-J/\psi$ invariant mass
distribution from $B\to K \pi^+\pi^-J/\psi$ decays. With the most
precise measurement of the mass of $M^{CDF}_{X(3872)} = 3871.61\pm
0.16\pm 0.19$~MeV~\cite{CDF_x3872_mass} from CDF, we know the mass
to be $3871.56\pm 0.22$~MeV~\cite{pdg2010}, which is very close to
the $D^{*0}\bar{D^0}$ mass threshold: $m_{D^{*0}}+m_{D^0} =
3871.78\pm 0.29$~MeV~\cite{pdg2010}. This suggests a binding
energy of $-0.22\pm 0.36$~MeV if $X(3872)$ is interpreted as a
$D^{*0}\bar{D^0}$ molecule, to be compared with the binding energy
of $-2.2$~MeV in deuteron case.

The quantum number of the $X(3872)$ was found to be either
$1^{++}$ or $2^{-+}$. A study of the $\pi^+\pi^-$ mass
distribution~\cite{CDF_jpc} and the observation of its $\gamma
J/\psi$ decays~\cite{belle_gpsi,babar_gammajpsi} indicate that the
$C$-parity of the $X(3872)$ is even, and the angular correlations
among the $\pi^+\pi^-J/\psi$ final state particles constrains the
$J^{PC}$ for the $X(3872)$ to be $1^{++}$ or $2^{-+}$, with
$1^{++}$ preferred~\cite{CDF_jpc}. Recently, the $2^{-+}$
assignment was revived by BaBar's analysis of the
$\pi^+\pi^-\pi^0$ mass distribution in $\pi^+\pi^-\pi^0J/\psi$
final state, where $2^{-+}$ is found to be slightly favored than
$1^{++}$~\cite{babar_x3872_omegajpsi,Arafat}.

BaBar reported >$3\sigma$ significance signals for $X(3872)$
decays to both $\gamma J/\psi$ and $\gamma\psi(2S)$, and the
branching fraction of $X(3872)\to \gamma \psi(2S)$ is found to be
larger than that of $X(3872)\to \gamma
J/\psi$~\cite{babar_gammajpsi}. While at Belle~\cite{Watson},
$X(3872)\to \gamma J/\psi$ was observed with similar production
rate as measured by BaBar, but no $X(3872)\to \gamma \psi(2S)$
signal was observed, and ${\cal B}(B^+\to X(3872)K^+)\times {\cal
B}(X(3872)\to \gamma \psi(2S))<3.4\times 10^{-6}$ at the 90\% C.L.
was determined, smaller than the BaBar's result of $(9.5\pm 2.7\pm
0.6)\times 10^{-6}$~\cite{babar_gammajpsi}.

BaBar sets an upper limit of the $X(3872)$ production rate in the
$B$-meson decays by measuring the momentum distribution of the
inclusive kaon from $B$-meson decays~\cite{PRL96_babar}: \( {\cal
B}(B^-\to K^- X(3872))<3.2\times 10^{-4} \) at the 90\% C.L.
Together with all the other measurements on the product branching
fractions ${\cal B}(B^-\to K^- X(3872))\cdot {\cal B}(X(3872)\to
exclusive)$, one gets \( 2.3\%<{\cal B}(X(3872)\to
\pi^+\pi^-J/\psi)<6.6\%, \) \( 1.4\times 10^{-4}<{\cal B}(B^-\to
K^- X(3872))<3.2\times 10^{-4}, \) at the 90\% C.L. We find that
the decay width of the $X(3872)$ to $\pi^+\pi^-J/\psi$ is larger
and the production rate of the $X(3872)$ in $B$ decays is smaller
than conventional charmonium states such as $\eta_c$, $\psi(2S)$,
and $\chi_{c1}$~\cite{pdg2010}.

\subsection{The $XYZ$ states near 3.94~GeV}

In 2005, Belle reported observations of three states with masses
near 3940~MeV: the $X(3940)$, seen as a $D^*\bar{D}$ mass peak in
$e^+e^-\to J/\psi D^*\bar{D}$ annihilations~\cite{belle_x3940};
the $Y(3940)$, seen as an $\omega J/\psi$ mass peak in $B\to
K\omega J/\psi$~\cite{belle_y3940}; and the $Z(3930)$, seen as a
$D\bar{D}$ mass peak in $\gamma\gamma\to
D\bar{D}$~\cite{belle_z3930}. The $Y(3940)$ and $Z(3940)$ were
confirmed by BaBar experiment~\cite{babar_y3940,babar_z3940}.

The $Y(3940)$ mass is well above $D\bar{D}$ or $D^*\bar{D}$ mass
threshold, but was discovered via its decay to the hidden charm
$\omega J/\psi$ final state. This implies an $\omega J/\psi$
partial width that is much larger than expectations for usual
charmonium. Recently, BaBar~\cite{babar_x3872_omegajpsi,Arafat}
reported a study of $B\to K\omega J/\psi$ in which the $\omega
J/\psi$ invariant mass distribution shows clear evidence for the
$X(3872)$ and $Y(3940)$. However, the BaBar values for mass and
width of the $Y(3940)$ are both lower than the corresponding
values reported by Belle and are more precise:
$M=3919.1^{+3.8}_{-3.5}\pm 2.0$~MeV (BaBar) compared to $3943\pm
11\pm 13$~MeV (Belle), and $\Gamma=31^{+10}_{-8}\pm 5$~MeV (BaBar)
compared to $87\pm 22\pm 26$~MeV (Belle). Part of the difference
might be attributable to the larger data sample used by BaBar
(426~fb$^{-1}$ compared to Belle's 253~fb$^{-1}$), which enabled
them to use smaller $\omega J/\psi$ mass bins as well as
considering $X(3872)$ and $Y(3940)$ simultaneously in their
analysis.

To add more information to the states in this mass region, Belle
observed a dramatic and rather narrow peak, $X(3915)$, in
$\gamma\gamma\to \omega J/\psi$~\cite{belle_y3915} that is
consistent with the mass and width reported for the $Y(3940)$ by
the BaBar group. The resonance parameters of the $X(3915)$ are \(
M = 3914\pm 4\pm 2~{\rm MeV} \) and $ \Gamma = 28\pm
12^{+2}_{-8}~{\rm MeV}$, with a statistical significance of
$7.1\sigma$.

Of these, only the $Z(3930)$ has been assigned to a $2^3P_2$
$c\bar{c}$ charmonium state, which is commonly called the
$\chi_{c2}(2P)$. If not exotic states, the $X(3940)$ could be a
candidate for $\eta_c(3S)$ and the $Y(3940)$ (maybe the same as
the $X(3915)$) could be $\chi_{c0}(2P)$.

\subsection{The $Y(4140)$, $Y(4280)$, and $X(4350)$}

Using exclusive $B^+ \to J/\psi \phi K^+$ decays, CDF
Collaboration observed a narrow structure, $Y(4140)$, near the
$J/\psi \phi$ mass threshold with a statistical significance of
3.8$\sigma$~\cite{CDF}. The analysis is updated with more data and
the signal significance is larger than $5\sigma$ now~\cite{yik}.
The mass and width of this structure are fitted to be
$4143.4^{+2.9}_{-3.0}(\rm stat)\pm 0.6(\rm syst)~\hbox{MeV}$ and
$15.3^{+10.4}_{-6.1}(\rm stat)\pm 2.5(\rm syst)~\hbox{MeV}$
respectively. Furthermore, there is another narrow structure,
$Y(4280)$, with a significance of $3.1\sigma$ observed with
$M=4274.4^{+10.4}_{-6.1}(\rm stat)\pm 1.9(\rm syst)~\hbox{MeV}$
and $\Gamma=32.3^{+21.9}_{-15.3} (\rm stat)\pm 7.6 (\rm
syst)~\hbox{MeV}$. They are isospin singlet states with positive
$C$ and $G$ parities since the quantum numbers of both $J/\psi$
and $\phi$ are $I^G(J^{PC})=0^-(1^{--})$. It was argued by the CDF
Collaboration that the $Y(4140)$ can not be a conventional
charmonium state, because a charmonium state with mass about
4143~$\hbox{MeV}$ would dominantly decay into open charm pairs,
and the branching fraction into the double OZI forbidden modes
$J/\psi \phi$ or $J/\psi \omega$ would be negligible.

The Belle Collaboration searched for this state using the same
process with $772\times 10^6$ $B \bar{B}$ pairs. No significant
signal was found, and the upper limit on the production rate
${\cal B}(B^+\to Y(4140)K^+$, $Y(4140)\to J/\psi \phi)$ is
measured to be $6\times 10^{-6}$ at the 90\% C.L. Although this
upper limit is lower than the central value of the CDF measurement
$(7.7\pm 3.7)\times 10^{-6}$, they do not contradict with each
other considering the large error~\cite{CDF-note}.

Assuming the $Y(4140)$ is a $D_{s}^{\ast+} {D}_{s}^{\ast-}$
molecule with quantum number $J^{PC}=0^{++}$ or $2^{++}$, the
authors of Ref.~\cite{tanja} predicted a two-photon partial width
of the $Y(4140)$ of the order of 1~keV, which is large and can be
tested with experimental data. The Belle Collaboration searched
for this state in two-photon process~\cite{x4350}. No $Y(4140)$
signal is observed, and the upper limit on the product of the
two-photon decay width and branching fraction of $Y(4140) \to \phi
J/\psi$ is measured to be $\Gamma_{\gamma \gamma}(Y(4140)) {\cal
B}(Y(4140)\to\phi J/\psi)<39~\hbox{eV}$ for $J^P=0^+$, or
$<5.7~\hbox{eV}$ for $J^P=2^+$ at the 90\% C.L. The upper limit is
lower than the predictions in Ref.~\cite{tanja}. This disfavors
the scenario of the $Y(4140)$ being a $D_{s}^{\ast+}
{D}_{s}^{\ast-}$ molecule with $J^{PC}=0^{++}$ or $2^{++}$.

Evidence is reported for a narrow structure at
$4.35~\hbox{GeV}/c^2$ in the $\phi J/\psi$ mass spectrum in the
above two-photon process $\gamma \gamma \to \phi J/\psi$ in Belle
experiment~\cite{x4350}. A signal of $8.8^{+4.2}_{-3.2}$ events,
with statistical significance of greater than 3.2 standard
deviations, is observed. The mass and natural width of the
structure (named as $X(4350)$) are measured to be
$4350.6^{+4.6}_{-5.1}(\rm{stat})\pm 0.7(\rm{syst})~\hbox{MeV}$ and
$13.3^{+17.9}_{-9.1}(\rm{stat})\pm 4.1(\rm{syst})~\hbox{MeV}$,
respectively. It is noted that the mass of this structure is well
consistent with the predicted values of a $c\bar{c}s\bar{s}$
tetraquark state with $J^{PC}=2^{++}$ in Ref.~\cite{stancu} and a
$D^{\ast+}_s {D}^{\ast-}_{s0}$ molecular state in
Ref.~\cite{zhangjr}.

\subsection{The $Y$ states in $ISR\,\,$ processes}

The study of charmonium states via initial state radiation ($ISR$)
at the $B$-factories has proven to be very fruitful. In the
process $e^+e^- \to \gamma_{ISR} \pi^+\pi^-J/\psi$, the BaBar
Collaboration observed the $Y(4260)$~\cite{babary}. This structure
was also observed by the CLEO~\cite{cleoy} and Belle
Collaborations~\cite{belley} with the same technique; moreover,
there is a broad structure near 4.008~GeV in the Belle data. In a
subsequent search for the $Y(4260)$ in the $e^+e^- \to
\gamma_{ISR} \pi^+\pi^-\psi(2S)$ process, BaBar found a structure
at around 4.32~GeV~\cite{babar_pppsp}, while the Belle
Collaboration observed two resonant structures at 4.36~GeV and
4.66~GeV~\cite{belle_pppsp}. Recently, CLEO collected
13.2~pb$^{-1}$ of data at $\sqrt{s}=4.26$~GeV and investigated 16
decay modes with charmonium or light hadrons~\cite{cleoy4260}. The
large $e^+e^- \to \pi^+\pi^-J/\psi$ cross section at this energy
is confirmed.

Belle and BaBar have exploited $ISR$ to make measurements of cross
sections for exclusive open-charm final states in this energy
range~\cite{pakhlova_ddstr,pakhlova_dd,
pakhlova_lclcbar,pakhlova_ddbarpi,pakhlova_dstrdbarpi,babar_dd}.
The exclusive channels that have been measured so far nearly
saturates the total inclusive cross section, but there is no
evidence for peaking near the masses of the $Y$ states. The one
exception is $e^+e^-\to \Lambda^+_c \Lambda_c^-$, which has a
threshold peak in the vicinity of the $Y(4660)$ peak
mass~\cite{pakhlova_lclcbar}.

The cross sections of charm strange meson pair production
($\sigma(e^+e^-\to D^+_s D^-_s)$, $\sigma(e^+e^-\to D^{+*}_s
D^-_s+c.c.)$ and $\sigma(e^+e^-\to D^{+*}_s D^{-*}_s)$) were
measured by BaBar~\cite{Izen} and Belle~\cite{pakhlova_dsds}
recently. However, the data show no indication of any of the $Y$
states, as in the non-strange charmed meson pair final states.

The absence of any evidence for the $Y(4260)$ ($Y(4360)$) decays
to open charm implies that the $\pi^+\pi^- J/\psi$ ($\pi^+\pi^-
\psi(2S)$) partial width is large: the analysis of
Ref.~\cite{moxh} gives a 90\% C.L. lower limit $\Gamma(Y(4260)\to
\pi^+\pi^- J/\psi)>508$~keV, which should be compared with the
corresponding $\pi^+\pi^- J/\psi$ partial widths of established
$1^{--}$ charmonium states: 201~keV for the $\psi(2S)$ and
52.7~keV for the $\psi(3770)$~\cite{pdg2010}.

The nature of these states is unknown, and many possibilities have
been proposed~\cite{qwg_review}, including charmonium, molecule,
hybrid, baryonium, and so on. There are also argument that these
structures are purely threshold effect or the final state
interaction of the charmed meson pairs. More experimental
measurements and more theoretical calculations are badly needed to
have a deeper understanding of these structures.

\subsection{Multiple solutions in fitting $R$-values~\cite{wym_psi}}

In explaining the $Y$ states and the excited $\psi$ states, the
leptonic partial widths provide very important information. As we
know, the vector quarkonium states could be either $S$-wave or
$D$-wave spin-triplet states, with the $S$-wave states couple
strongly to lepton pair while the $D$-wave states couple weakly
since the latter are only proportional to the second derivative of
the wave-function at the origin squared, as expected in the
potential models. This leads people to believe that the
$\psi(4040)$ is the $3S$ charmonium state, $\psi(4160)$ the $2D$
state, and $\psi(4415)$ the $4S$ state. This has been a well
accepted picture for more than two decades before the discovery of
the so-called $Y$ particles, namely, the $Y(4008)$, $Y(4260)$,
$Y(4360)$, and $Y(4660)$. With seven states observed between
4.0~GeV/$c^2$ and 4.7~GeV/$c^2$, some people started to categorize
some of these as non-conventional quarkonium states, while others
tried to accommodate all of them in modified potential models.
Many of the theoretical models use the leptonic partial widths of
these states to distinguish them between $S$- and $D$-wave
assumptions~\cite{Klempt,KTChao}, and most of the time, the values
on the leptonic partial widths are cited from the
PDG~\cite{pdg2010} directly. Although the resonance parameters of
these excited $\psi$ states have been measured by many
experimental groups, all of them were obtained by fitting the
$R$-values measured in the relevant energy region. The most recent
ones~\cite{bespsires}, which were from a sophisticated fit to the
most precise $R$-values measured by the BES
collaboration~\cite{besr1,bes_Rhad}, are the only source of the
leptonic partial widths of these three $\psi$ states now quoted by
the PDG~\cite{pdg2010}.

In fitting to the BES data, unlike the previous analyses, the BES
collaboration considered the interference between the three
resonances decaying into the same final modes, and introduced a
free relative phase for the amplitude of each
resonance~\cite{bespsires}. The new parametrization of the
hadronic cross section results in a pronounced increase of the
$\psi(4160)$ mass, and significant decrease of the leptonic
partial widths of $\psi(4160)$ and $\psi(4415)$.

In a recent study~\cite{wym_psi}, it is pointed out multiple
solutions can be found in the fit to the $R$-values, resulting
four sets of solutions for the leptonic partial widths of the
excited $\psi$ states. The fit results are presented in
Table~\ref{tab_datfit}, there are four solutions found in the fit.
It should be noted that the four solutions have identical
$\chi^2$, masses, and total widths for the resonances, but
different partial widths to lepton pairs.

\begin{table}
\caption{\label{tab_datfit} Four groups of solutions for the data
fitting. The four solutions have identical resonance masses ($M$)
and total widths ($\Gamma_t$), but significant different leptonic
partial widths ($\Gamma_{ee}$) and the relative phases ($\phi$). }
\begin{center}
\begin{tabular}{cccc}
\hline\hline
  Parameter        & $\psi(4040)$ & $\psi(4160)$   & $\psi(4415)$ \\
\hline
$M$ (MeV)       &$4034\pm  6$  &$4193\pm  7$  &$4412\pm 15$ \\
$\Gamma_t$ (MeV) &$  87\pm 11$  &$  79\pm 14$  &$ 118\pm 32$
\\\hline\hline
 $\Gamma^{(1)}_{ee}$ (keV)
           &~$0.66\pm 0.22$~&~$0.42\pm 0.16$~&~$0.45\pm 0.13$~ \\
$\phi^{(1)}$ (radian)
           & 0 (fixed)    &$2.7\pm 0.8$  &$2.0\pm 0.9$ \\\hline
$\Gamma^{(2)}_{ee}$ (keV)
           &$0.72\pm 0.24$&$0.73\pm 0.18$&$0.60\pm 0.25$ \\
$\phi^{(2)}$ (radian)
           & 0 (fixed)    &$3.1\pm 0.7$  &$1.4\pm 1.2$ \\\hline
$\Gamma^{(3)}_{ee}$ (keV)
           &$1.28\pm 0.45$&$0.62\pm 0.30$&$0.59\pm 0.20$ \\
$\phi^{(3)}$ (radian)
           & 0 (fixed)    &$3.7\pm 0.4$  &$3.8\pm 0.8$ \\\hline
$\Gamma^{(4)}_{ee}$ (keV)
           &$1.41\pm 0.12$&$1.10\pm 0.15$&$0.78\pm 0.17$ \\
$\phi^{(4)}$ (radian)
           & 0 (fixed)    &$4.1\pm 0.1$  &$3.2\pm 0.3$
           \\\hline\hline
\end{tabular}
\end{center}
\end{table}

As the $\Gamma_{ee}$ of the vector resonances are closely related
to the nature of these states, the choice among the distinctive
solutions affects the classification of the charmonium and
charmonium-like states observed in this energy region. The
$Y(4260)$ was proposed to be the $\psi(4S)$ state and the
$\psi(4415)$ be $\psi(5S)$ in Refs.~\cite{Klempt,KTChao}, we can
see that in this assignment, the calculated partial widths of
$\psi(4040)=\psi(3S)$ and $\psi(4415)=\psi(5S)$~\cite{KTChao}
agree well with the fourth solution listed in
Table~\ref{tab_datfit}.

Of course the possible mixing between $S$- and $D$-wave states
will change significantly the theoretical predictions of the
partial widths of these states~\cite{badalian}, and the QCD
correction, which is not well handled~\cite{wanggl}, may also
change the theoretical predictions significantly. So far, we have
no concrete criteria to choose any one of the solutions as the
physics one.

It should be noticed that if the $Y$ states are considered
together with the excited $\psi$ states in fitting the $R$-values,
there could be even more solutions, the situation may become more
complicated. We also notice that the existence of the multiple
solutions is due to the inclusion of a free phase between two
resonances, if these phases can be determined by other means
(either theoretically or experimentally), it will be very helpful
to know which solution corresponds to the real physics.

\section{Summary} 

In summary, there are lots of progress in low and high mass
spectroscopy. A better understanding of the charmonium and
bottomonium spin-singlets is achieved, although $h_b$ is still not
observed. There are lots of new states, composed of either only
light quarks or light and heavy quarks, observed in experiments.
Some of these states have very exotic properties, which may
suggest the long searching exotic states have been observed.
However, due to limited statistics, the experimental information
on the properties of any of these states is not enough for us to
draw solid conclusion, let alone our poor knowledge on the QCD
prediction of the properties of the exotic states or the
conventional mesons.

In the near future, BESIII experiment~\cite{bes3} may accumulate
more data for center of mass energy between 3.0 and 4.6~GeV, this
will contribute to the understanding of some of the states
discussed above; the Belle II experiment~\cite{belle2} under
construction, with about 50~ab$^{-1}$ data accumulated, will
surely improve our understanding of all these states. We also
expect new results from hadron colliders as well as fixed target
experiments.

\section*{Acknowledgment}

We thank the organizers for their kind invitation and congratulate
them for a successful conference. This work is supported in part
by National Natural Science Foundation of China under Contract
Nos. 10825524 and 10935008.

\end{document}